# OPTIMIZATION FOR GROWTH CONDITION OF ULTRATHIN HEXAGONAL BORON NITRIDE ON DIELECTRIC SUBSTRATES VIA LPCVD METHOD

Meryem Bozkaya[1,*], Muhammet Nasuh Arık[2], Ali Altuntepe[3], Hakan Ateş[4], Recep ZAN[5]

[1] Gazi University, Graduate School of Natural and Applied Science, Advanced Technologies, Ankara, TÜRKİYE.
[2] Turkish Energy, Nuclear and Mineral Research Agency (TENMAK) Boron Research Institute, Ankara, TÜRKİYE.
[3] Sivas University of Science and Technology, Optical Excellence Application and Research Center, Sivas, TÜRKİYE.
[4] Gazi University, Faculty of Technology, Metallurgical and Materials Engineering, Ankara, TÜRKİYE.
[5] Niğde Ömer Halisdemir University, Faculty of Science, Physics Department, Niğde, TÜRKİYE.
*Corresponding Author
ORCID1: 0000-0001-5711-9369, ORCID2: 0000-0002-1526-803X, ORCID3: 0000-0002-6366-4125, ORCID4: 0000-0002-5132-4107, ORCID5: 0000-0001-6739-4348
meryemm.bozkaya@gmail.com , nasuh.arik@tenmak.gov.tr , alialtuntepe@sivas.edu.tr , hates@gazi.edu.tr , recep.zan@ohu.edu.tr

**Özet**

Hekzagonal Bor Nitrür (h-BN) katman sayısına bağlı olarak değişen geniş bant aralığı, yüksek mekanik ve termal özellikleri, yüksek kenar soğurma katsayısı ve ultraviyole (UV) bölgede saydamlık gibi benzersiz nitelikleri sayesinde; derin UV fotodedektörler, UV gözlem ve iletişim sistemleri ile güneş hücreleri gibi heteroyapılı optoelektronik cihaz uygulamaları için oldukça ilgi çekici bir adaydır. H-BN ince filmlerin bu uygulamalarda kullanımı için yaygın olarak kullanılan yöntem Kimyasal Buhar Biriktirme (CVD) yöntemidir ve genellikle sentez esnasında bakır (Cu) ve Nikel (Ni) gibi alttaşlar tercih edilmektedir. Ancak, uygulama alanına entegrasyon için bu katalitik alttaşlardan hedef dielektrik alttaşlara transfer edilmesi gerekmektedir. Transfer süreci, malzemede katlanmalar, çatlaklar ve polimer kalıntıları gibi kusurlara yol açarak malzemenin optoelektronik özelliklerini bozmakta ve cihaz performansını ciddi ölçüde düşürmektedir. Bu sorunun üstesinden gelmek amacıyla, h-BN filmlerin silikon (Si), $SiO_2$, kuvars, safir veya AlN gibi dielektrik alttaşlar üzerine doğrudan (transfer gerekmeden) sentezlenmesi hedeflenmektedir. Ancak, katalitik etki eksikliği, malzemenin homojen ve yüksek kristallikte ve kontrol edilebilir kalınlıkta sentezlenmesinde önemli zorluklar doğurmaktadır. Bu çalışmada, UV bölgede yüksek saydamlığa sahip katalitik olmayan kuvars alttaşlar üzerinde ultra ince h-BN filmlerin düşük basınçlı kimyasal buhar biriktirme (LPCVD) yöntemiyle doğrudan sentezlenmesi için büyütme parametrelerinin optimizasyonu incelenmiştir. Yapılan çalışmalar sonucunda, 1050°C'de 60 dak süre ile 150 mg Amonyak Boran (AB) öncül maddesinin 80°C'de bozunmasıyla edildiği belirlenmiştir.

# Optimization For Growth Condition of Ultrathin Hexagonal Boron Nitride on Dielectric Substrates via LPCVD Method

## Abstract

Hexagonal Boron Nitride (h-BN) is a highly intriguing candidate for heterostructure optoelectronic applications, such as Deep Ultraviolet photodetectors, UV sensing and communication systems and solar cells. This is primarily due to its unique properties, including a layer dependent wide energy bandgap, superior mechanical strength, high thermal conductivity, high band-edge absorption coefficient, and exceptional transparency in the UV region. The widely adopted synthesis method for h-BN thin films is Chemical Vapor Deposition (CVD) Method, which often utilizes catalytic substrates like copper (Cu) and Nickel (Ni). However, integrating the synthesized h-BN into device applications requires a subsequent transfer process to the target substrate. This transfer step introduces significant material damage, such as folding, cracking and polymer residues, which ultimately degrade the optoelectronic properties of the material and compromise device performance. To overcome this major challenge, there is a strong need to synthesize high- quality h-BN films directly onto dielectric substrates such as silicon (Si), $SiO_2$, quartz, sapphire or AlN without the need for transfer. The primary difficulty in direct synthesis lies in achieving homogenous, high crystallinity films with controllable thickness due to the absence of a catalytic effect. In this work, we investigated the optimization of growth parameters for the direct synthesis of ultrathin h-BN films on non-catalytic quartz substrates, which are highly transparent in the UV region, using the Low-pressure Chemical Vapor Deposition (LPCVD) method. The optimal synthesis conditions were determined to be 1050°C for 60 min, achieved by the decomposition of 150 mg Ammonia Borane (AB) precursor at 80°C. This optimization is crucial for advancing large-scale, high-performance h-BN based DUV photodetector fabrication.

1. Introduction

The development of optoelectronic devices operating in the Deep Ultraviolet (DUV) region [1] (∼200–280 nm) is a crucial technological frontier with significant applications ranging from military surveillance and missile warning systems [2] (MWS) to flame detection, non-line-of-sight (NLOS) optical communications [3], and biological agent sensing [4]. Devices targeting this range—known as the "solar-blind" region—offer a superior signal-to-noise ratio since natural solar radiation at these wavelengths is efficiently absorbed by the Earth's atmosphere's ozone layer [5]. Conventional wide bandgap semiconductors like Gallium

Nitride (GaN) [6], Silicon Carbide (SiC) [7], and Gallium Oxide ($Ga_2O_3$) [8] have long been the primary materials for DUV photodetector fabrication [6-8]. However, the emergence of two-dimensional (2D) materials offers opportunities to address limitations such as high defect density, high processing costs, and the complexity associated with conventional bulk semiconductor fabrication [8,9]. Among the 2D materials, Hexagonal Boron Nitride (h-BN) stands out as an exceptional candidate for next-generation DUV applications [9, 10]. h-BN is a structural analog of graphene, possessing an ultra-wide bandgap (UWBG) typically around 6.0–6.4 eV [11], which naturally positions it to operate precisely within the DUV spectral window [12,13]. Furthermore, h-BN exhibits outstanding physical and chemical properties, including superior mechanical strength, high thermal stability and conductivity, and a high absorption coefficient at the band edge ($7,5 \times 10^5$ $cm^{-1}$) [14]. These combined attributes make h-BN highly desirable for creating robust, highly sensitive, and thermally stable photodetectors [15,16]. For large-scale and high-quality synthesis, the Chemical Vapor Deposition (CVD) method remains the most scalable and controllable technique for h-BN [17]. However, traditional CVD processes necessitate the use of catalytic metal substrates, such as copper (Cu) [18] or nickel (Ni) [19], to facilitate the growth. This approach inevitably introduces a critical technological hurdle: the resulting h-BN film must be transferred from the metallic catalyst onto the final target substrate (e.g., $SiO_2$ or quartz) for device integration [20,21].

The transfer process is highly detrimental to the material's integrity [21]. It often leads to the incorporation of significant defects, including tears, folds, cracks, and chemical residues from the wet etching or polymer-assisted removal steps [21,22]. These structural and chemical contaminations dramatically impair the material's crystalline quality and electronic properties, resulting in low reproducibility and compromised device performance—a major barrier to the commercialization of h-BN-based optoelectronics [23]. To circumvent the inherent drawbacks of the transfer process, the scientific community has been actively pursuing the direct synthesis of h-BN thin films onto dielectric or optically transparent substrates (Si, $SiO_2$, sapphire, quartz) where devices can be fabricated directly [23-26]. Despite the significant advantages, direct synthesis is challenging. The absence of a strong catalytic effect makes it difficult to achieve large-area uniformity, high crystallinity, and precise control over the film thickness, particularly in the ultra-thin regime necessary for high-performance 2D devices [27-28].

This study addresses this critical challenge by focusing on the parameter optimization of the Low-Pressure Chemical Vapor Deposition (LPCVD) method for the direct synthesis of ultra-thin h-BN films on non-catalytic quartz substrates. Quartz was chosen for its excellent transparency in the target UV region, making it ideal for DUV photodetector applications. The primary goal of this work is to systematically investigate the correlation between key LPCVD growth parameters—namely growth temperature, time, and precursor concentration (Ammonia Borane) —and the resulting structural and optical quality of the h-BN film. This paper is organized as follows: Section II details the experimental methodology and the LPCVD setup. Section III presents the comprehensive characterization results, elucidating the effects of the optimized parameters. Section IV discusses the potential of the optimized h-BN films for DUV photodetector applications, followed by the main conclusions in Section V. We successfully determined the optimal synthesis conditions to be 1050°C for 60 minutes with 150 mg of precursor decomposed at 80°C, providing a scalable and contamination-free pathway for high-performance h-BN based optoelectronics.

## 2. Materials and Methods

### 2.1. Materials

The materials used for the synthesis and characterization of h-BN films were carefully selected and sourced. Ammonia Borane (AB) which served as the precursor for Boron and Nitrogen, was obtained with 97% purity from the Boron Research Institute (BOREN) of the Turkish Energy, Nuclear and Mineral Research Agency (TENMAK). Additionally, the quartz substrates (1x1,5 cm$^2$) used for film growth were purchased from Süber Laboratuvar Cam Laboratuvar Ürünleri ve Kimyasalları Company.

### 2.2. Methods

The Chemical Vapor Deposition (CVD) method is widely used for the synthesis of ultra-thin hexagonal Boron Nitride (h-BN) films due to its environmentally friendly nature, suitability for large-scale production, tunability, and relatively low cost [17]. The setup used for the Low-Pressure CVD (LPCVD) system is shown schematically in Figure 1.

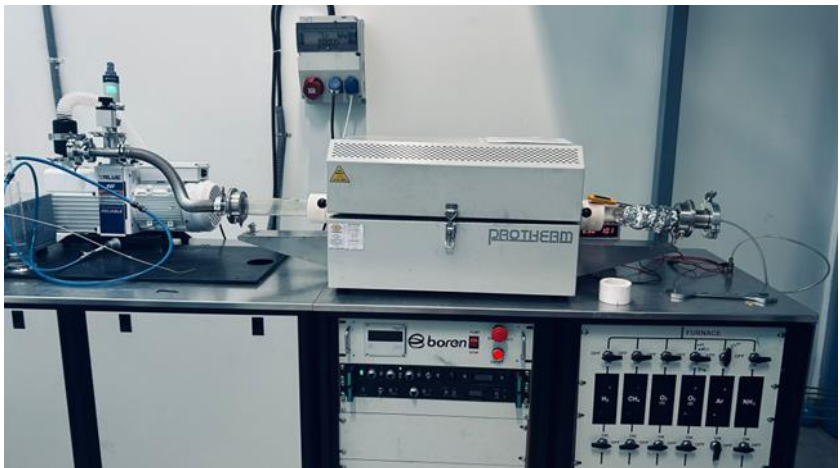

Fig.1. LPCVD system for the h-BN growth.

First, 1x1,5cm$^{-2}$ quartz substrates were cleaned in an ultrasonic bath sequentially with acetone, isopropyl alcohol (IPA), and deionized water (DIW) for 15 minutes each. The cleaned substrates were dried with compressed air and swiftly loaded into a 120-cm-long quartz tube with a 20 mm inner diameter. Subsequently, the Ammonia Borane (AB) precursor, contained in an alumina boat, was placed 50 cm away from the substrates, outside the main furnace, and heated using a heating belt. Prior to initiating the furnace, the system was subjected to two cycles of Argon (Ar) purging to ensure the removal of residual oxygen. The furnace was then ramped up to 1050ºC at a rate of 10ºC/min under a flow of 300 sccm of 10% Ar/H$_2$ (Argohid). As soon as the furnace reached the required temperature, the decomposition of the AB precursor was initiated via the heating belt. The system pressure prior to AB precursor decomposition was 6,50x10$^{-2}$ mbar. A corresponding increase in partial pressure occurred during decomposition, depending on the amount of AB used. The following ranges were determined for the synthesis parameters (Table 1) growth temperature (1000-1050ºC), growth duration (90–15 min), precursor amount (200–50 mg), and precursor decomposition temperature (80-100ºC). Following the growth, the system was allowed to cool under a continuous flow of 500 sccm of 10% Ar/H$_2$ (Argohid) gas.

Table.1. Recipe of h-BN thin film synthesis.

| Temperature | **1000-1050°C** |
|---|---|
| **Heating Rate** | 10°C/min. |
| **Substrate used** | Quartz |
| **Type of carrier gas** | %10 $H_2$ Argohid |
| **Amount of carrier gas** | 300sccm (270 sccm Ar; 30 sccm $H_2$) |
| **Type and quantity of precursors** | Amonyak Boran 50-200 mg |
| **Precursor Decomposition Temperature** | 80-100 °C |
| **Pressure** | $6,50 \times 10^{-1}$ mbar |
| **Growth Time** | 15-90 min. |

### 3. Results and Discussion

*Effect of Growth Time*

The chemical composition and bonding state of the h-BN films as a function of growth time were meticulously investigated using X-Ray Photoelectron Spectroscopy (XPS). The B1s and N1s core-level spectra (Figures 2a and 2b) consistently confirm the synthesis of the h-BN phase, indicated by the dominant, symmetric peaks centered at approximately **191.02 eV (B1s) and 398.52 eV (N1s)**, which are characteristic of $sp^2$-bonded Boron and Nitrogen atoms, respectively [29]. Quantitative analysis of the optimal 60 min. film (**detailed in Table 2**) confirmed high stoichiometry and chemical homogeneity, with the B1s and N1s atomic percentages recorded at 16.62% and 16.27% respectively, resulting in a near-ideal B: N ratio of 1.02. The high structural quality was further validated by the narrow FWHM values of 2.98 eV (B1s) and 2.28 eV (N1s), confirming a highly uniform $sp^2$ bonding environment. Crucially, the intensity of both the B1s and N1s peaks increases significantly as the growth time is extended from 15 min up to 90 min given that the synthesis occurs on an inert dielectric substrate (quartz), which follows a non-catalytic growth mechanism, this trend demonstrates that the film thickness is directly controllable by the deposition time. The continuous increase in intensity, even at 90 minutes, suggests that the growth process has not reached full saturation; However, the data confirms that 60 min provides the optimal balance for achieving high-purity, thickness-tunable layered h-BN film directly on the dielectric surface.

Table.2. XPS Peak Table of grown sample at 60min.

| Name | Start BE | Peak BE | End BE | FWHM eV | Atomic % |
|---|---|---|---|---|---|
| **Si2p** | 109,98 | 107,03 | 103,28 | 2,4 | 11,47 |
| **C1s** | 290,58 | 285 | 281,88 | 2,54 | 11,21 |
| **B1s** | 199,98 | 191,02 | 187,78 | 2,98 | 16,62 |
| **N1s** | 403,08 | 398,52 | 394,88 | 2,28 | 16,27 |
| **O1s** | 537,48 | 532,91 | 528,88 | 2,29 | 22,21 |

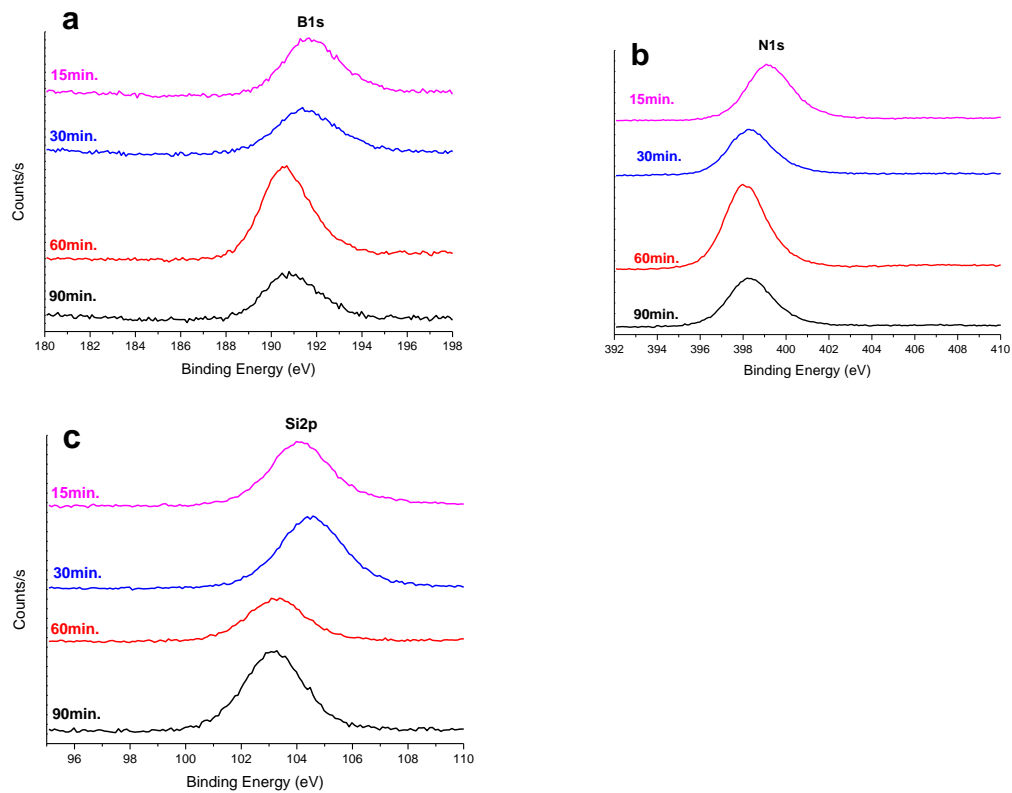

Fig.2. XPS analysis of samples grown at 1050C for 90-60-30-15 min. a) B1s b)N1s c) Si2p.

This interpretation is further supported by the **Si2p core-level spectra (Figure 2c)**, which show a notable **attenuation in intensity** as the growth time increases from 15 min to 90 min given that the Si2p peak originates from the underlying quartz ($SiO_2$) substrate, this clear reduction in signal confirms that the growing h-BN film acts as an **effective overlayer**, successfully covering and gradually increasing in thickness on the dielectric substrate [30].

*The Effect of AB Precursor Amount*

The influence of the AB precursor mass on the synthesized h-BN film was investigated while keeping the growth temperature and duration constant at 1050ºC and 60 minutes. The B1s and N1s core-level spectra (Figures 3a and 3b) consistently demonstrate that the characteristic $sp^2$ bonding energy remains stable, indicating that the AB amount does not alter the chemical phase of the product. **Quantitative analysis of the optimized 150mg film confirmed high chemical purity and ideal stoichiometry (Table 3).** The resulting B: N ratio of 1.03, along with the narrow FWHM values (2.49 eV for B1s and 2.27 eV for N1s), is highly desirable for high-quality h-BN films and demonstrates that the 150 mg mass provides sufficient precursor without promoting excessive B-rich defects. However, a significant observation is the proportional increase in the intensity of both the B1s and N1s signals as the precursor mass is increased from 50 mg to 200 mg. This relationship confirms that the h-BN synthesis is operating under a precursor-limited kinetic regime [21]. In this regime, increasing the solid AB mass leads to a higher concentration of active Boron and Nitrogen species in the gas phase, directly translating to a higher growth rate and a thicker h-BN film yield within the fixed growth time. Therefore, the AB precursor amount is established as a key parameter for achieving precise control over the final film thickness during the direct LPCVD synthesis on dielectric substrates.

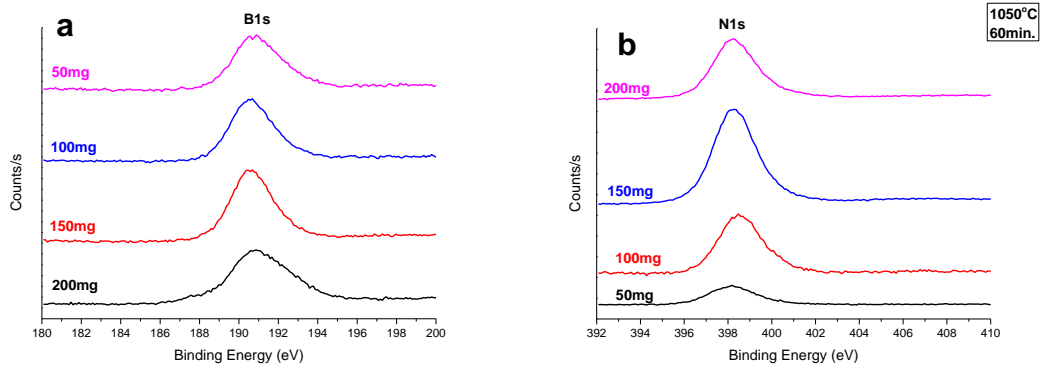

Fig.3. XPS analysis of samples grown at 1050C for 200-150-100-50 mg AB precursor a) B1s b) N1s.

Table3. XPS Peak Table of Grown sample with 150mg AB.

| Name | Start BE | Peak BE | End BE | FWHM eV | Atomic % |
|---|---|---|---|---|---|
| Si2p | 107,18 | 103,17 | 98,78 | 2,36 | 11,3 |
| B1s | 195,48 | 190,58 | 185,88 | 2,49 | 32,36 |
| C1s | 290,78 | 284,75 | 280,58 | 2,66 | 8,94 |
| N1s | 402,88 | 398,26 | 394,08 | 2,27 | 31,37 |
| O1s | 537,98 | 532,67 | 528,18 | 2,23 | 16,03 |

*The Effect of precursor decomposition Temperature*

The influence of the AB precursor decomposition temperature ($T_{dec}$) on the resulting h-BN film was investigated while fixing the synthesis temperature and duration (1050°C for 60 min) The B1s and N1s core-level spectra (Figures 4a and 4b) consistently show that the sp$^2$ bonding energy remains unchanged, validating that the chemical phase of the product is stable h-BN. Quantitative analysis of the film synthesized at the optimal 80°C $T_{dec}$ confirmed outstanding stoichiometry and chemical homogeneity (Table 4). The B1s and N1s atomic percentages, recorded at 33.83% and 33.34% respectively, yielded a near-perfect B:N ratio of 1.01, affirming high film purity. Furthermore, the exceptionally narrow and identical FWHM values for both B1s and N1s peaks (1.88 eV) demonstrate a highly uniform bonding environment, suggesting superior crystalline quality under these controlled conditions. However, a notable difference is observed in the peak intensity: the film synthesized with a $T_{dec}$ of 80°C **exhibits a higher signal intensity** compared to the film grown using a $T_{dec}$ of 100°C. The precursor decomposition temperature is a critical factor that controls the **flux (or partial pressure)** of active B-N species delivered to the reactor. In the non-catalytic LPCVD growth regime employed here, the lower decomposition temperature 80°C leads to a **slower, more controlled, and sustained release** of the precursor. This controlled flux is hypothesized to promote more efficient surface adsorption, improved lateral growth, and ultimately results in a **thicker and/or more uniform h-BN film coverage** than the rapid and potentially inefficient delivery achieved at 100°C. This finding establishes the precursor decomposition temperature as a key factor for optimizing film yield during direct synthesis on dielectric substrates.

Table.4. XPS Peak Table of Grown Sample at 80°C

| Name | Start BE | Peak BE | End BE | FWHM eV | Atomic % |
|---|---|---|---|---|---|
| Si2p | 107,38 | 103,43 | 99,48 | 2,18 | 8,89 |
| B1s | 195,58 | 190,65 | 186,28 | 1,88 | 33,83 |
| C1s | 291,68 | 284,89 | 280,98 | 2,13 | 10,16 |
| N1s | 403,18 | 398,19 | 394,08 | 1,88 | 33,34 |
| O1s | 538,08 | 532,67 | 527,78 | 2,1 | 13,77 |

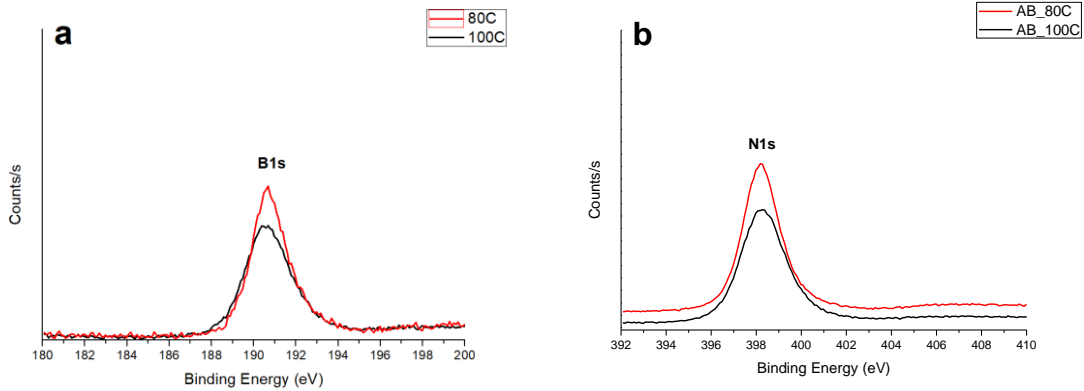

Fig.4. XPS analysis of samples grown at 1050ºC for 100C and 80ºC AB decomposition temperature a) B1s b) N1s.

*Photoluminescence (PL) Analysis of Temperature Dependence*

The optical quality and presence of deep-level defects in the h-BN films synthesized at 1000ºC and 1050ºC were investigated using Photoluminescence (PL) spectroscopy with a 390 nm excitation wavelength. Both spectra (Figure 5) display prominent, broad-band emission peaks in the visible region, specifically centered around 480nm, 570 nm and 610 nm [31-34]. These features are highly characteristic of **deep-level defect emissions** and are frequently observed in h-BN materials grown via CVD [35]. While the intrinsic bandgap emission of bulk h-BN is in the deep ultraviolet (200-200 nm) these visible emissions are generally attributed to structural defects such as **Boron vacancies ($V_B$)** nitrogen vacancies ($V_N$) or oxygen-related complexes, which form defect states within the bandgap [36].

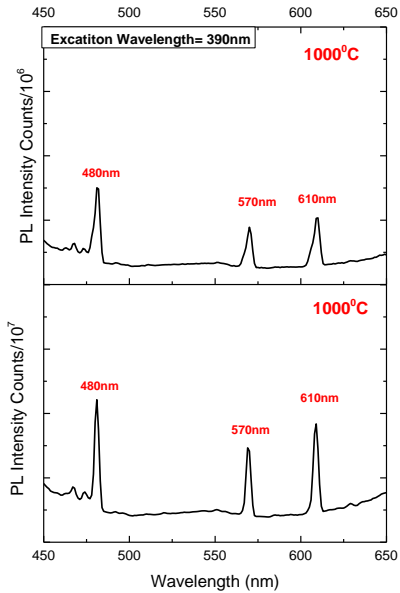

Fig.5. PL spectra of grown samples at 1050°C and 1000°C.

Comparing the two synthesis temperatures, the film grown at 1050°C exhibits a **higher overall PL intensity** across all observed defect peaks compared to the film grown at 1000°C. This indicates that, under the given LPCVD conditions, the higher growth temperature either resulted in a **thicker layer of h-BN (more emitting material)** or promoted the formation of a **higher concentration of the specific defects** responsible for these radiative transitions [36]. Given the synthesis goal is maximizing film yield on dielectric substrates, the enhanced PL signal at 1050°C suggests this temperature is more effective for establishing the desired h-BN structure or thickness.

*FT-IR Analysis of Precursor Amount Effect*

Fourier-Transform Infrared Spectroscopy (FT-IR) was employed to confirm the bonding structure and chemical purity of the h-BN films synthesized using 150 mg and 200 mg of AB precursor. Both spectra (Figure 6) display the two characteristic absorption bands of $sp^2$-bonded h-BN a sharp, intense peak centered at approximately **1375 cm$^{-1}$**, corresponding to the **in-plane B-N stretching mode** ($E_{1u}$), and a broader peak at approximately **780 cm$^{-1}$**, corresponding to the **out-of-plane B-N-B bending mode** ($A_{2u}$) [37]. The presence of these two peaks strongly confirms the successful formation of the layered hexagonal phase.

However, a critical difference is observed in the impurity signature: the film grown with the higher amount of precursor 200mg exhibits a distinct absorption feature at **1132cm$^{-1}$**. This peak is universally attributed in literature to the presence of **B-O stretching vibrations**, indicating the incorporation of oxygen or the formation of an undesirable $B_2O_3$ byproduct within the film or on its surface [38]. The presence of this impurity suggests that while higher AB mass (as supported by XPS data) leads to a thicker film, using **200mg under these LPCVD conditions promotes greater oxygen incorporation**. This likely stems from a combination of excess precursor species reacting with residual $H_2O$ or oxygen in the reactor, suggesting that the 150 mg precursor amount provides a better balance between film yield and chemical purity.

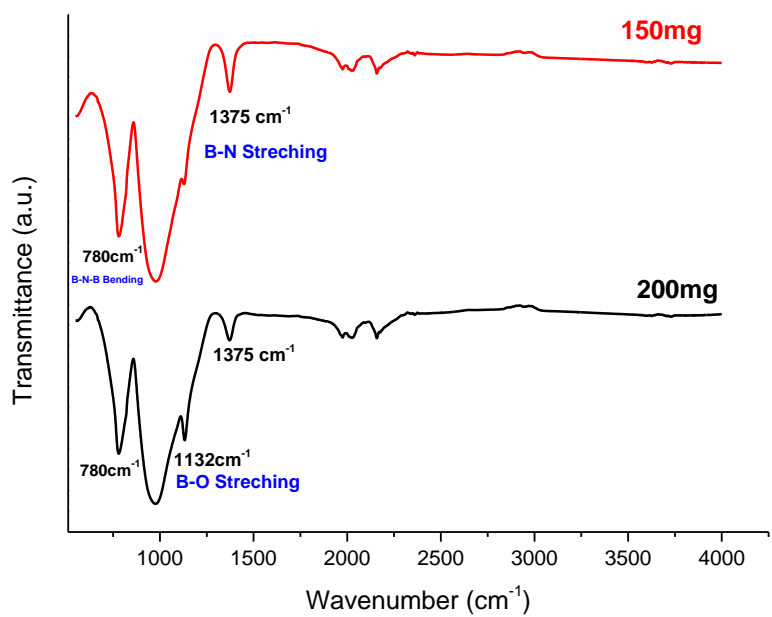

Fig.6. FT-IR spectra of grown sample at different AB precursor amount 200 and 150mg.

*Structural and Morphological Analysis*

Raman Spectroscopy was employed to confirm the structural quality and $sp^2$ phase of the h-BN film grown at 1050°C. The spectrum (Figure 7a) exhibits a sharp, intense characteristic peak at **1371cm$^{-1}$** which is the definitive signature of the in-plane $E_{2g}$ phonon mode, validating the formation of the layered $sp^2$-bonded h-BN lattice [39]. However, secondary features were also observed, including a peak at **1031cm$^{-1}$** attributed to B-O stretching vibrations, reflecting oxygen incorporation or residual $B_2O_3$ byproduct [40]. This structural characterization is complemented by the optical microscopy image (Figure 7b), which confirms the film's **excellent, near-complete coverage** and **highly uniform surface morphology** across the large area of the quartz substrate. This visual uniformity strengthens the conclusions drawn from the XPS attenuation data. The scattered, small dark spots visible in the optical image are consistent with typical CVD growth, likely representing **thicker, multi-layer h-BN clusters** or minor residual defects. The successful formation of a large-area, morphologically uniform film without visible cracking on a dielectric substrate confirms the efficacy of the optimized LPCVD process.

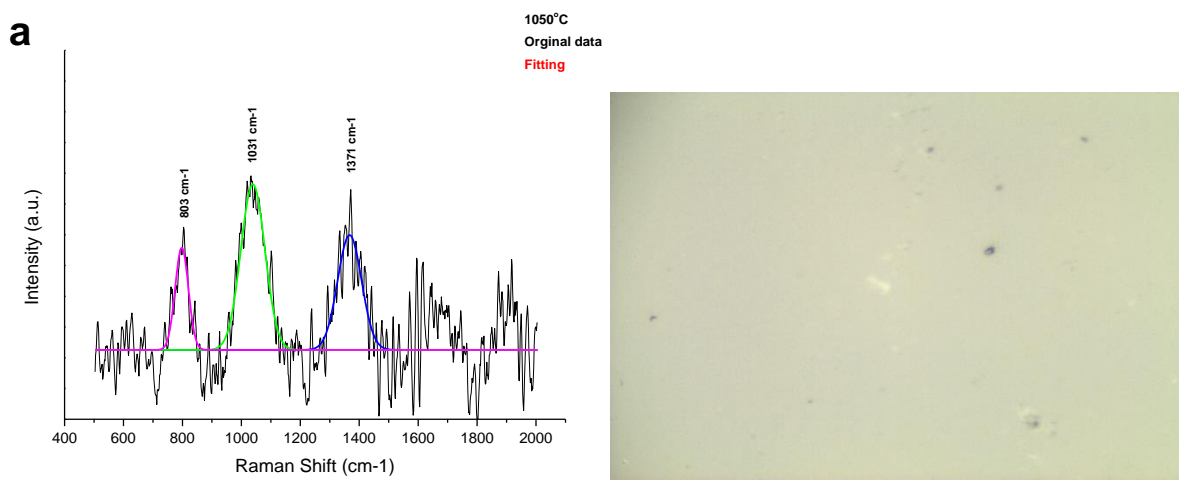

Fig.7. a) Raman Spectra of grown h-BN thin film b) Optic microscopy image of grown h-BN thin film on quartz substrate.

The comparative quantitative XPS analysis across the three optimized parameter conditions reveals a critical kinetic effect on film growth, thickness, and purity. While all films maintained excellent stoichiometry (B: N ratios between 1.01 and 1.03), the total h-BN signal varied dramatically, with the 80°C $T_{dec}$ and 150mg AB films showing nearly double the combined B and N signal approximately 67.17% and approximately 63.73%, respectively) compared to the 60 min growth time sample (approximately 32.89%).

This significant increase in the h-BN signal confirms that the 80°C $T_{dec}$ and 150mg AB parameters effectively break the bottleneck of the non-catalytic growth regime. The kinetic advantage is fully demonstrated by the substrate signal attenuation: the 80°C $T_{dec}$ film minimizes the Si2p signal from the underlying quartz to its lowest level (8.89%), validating that this condition yields the **highest film thickness and most complete coverage**.

Furthermore, this optimal kinetic control simultaneously resulted in the highest film quality: the 80°C $T_{dec}$ sample registered the **lowest O impurity signal** (13.77% O1s atomic percentage) and the **narrowest FWHM** (1.88eV) for both B1s and N1s peaks. This suggests that the slow, controlled precursor flux at 80°C not only maximizes the growth rate but also minimizes the incorporation of oxygen-related defects (as corroborated by the FT-IR analysis), yielding the most uniform and chemically pure h-BN film across all tested parameters.

**Conclusions**

This study successfully demonstrated the synthesis of high-quality h-BN thin films directly onto dielectric quartz substrates via LPCVD method, effectively eliminating the need for the critical and defect-inducing transfer process. Spectroscopic analysis confirmed the formation of the desired $sp^2$-bonded h-BN phase (validated by the characteristic $E_{2g}$ Raman peak and B1s /N1s XPS binding energies). The growth kinetics were established to be precursor-limited, with film thickness being precisely tunable through two critical parameters: growth duration (evidenced by the increase in B1s/N1s signal and simultaneous attenuation of the Si2p substrate signal) and Ammonia Borane (AB) precursor amount. Optimization showed that a 1050°C growth temperature yielded optimal film development, while a lower precursor decomposition temperature of 80°C provided a superior, controlled flux for maximizing film yield. Morphological characterization revealed excellent, large-area coverage and high uniformity on the quartz surface, supporting the film's mechanical integrity. Although the presence of B-O related impurities was identified by both FT-IR and Raman analyses, this work validates a robust, scalable, and direct growth strategy for producing thickness-controllable h-BN films, paving the way for advanced DUV device applications.

**Author Contribution Statement**

The individual contributions of the authors to this work are defined as follows:

**Meryem B.:** Was responsible for the Conceptualization, design of the Methodology, execution of the (LPCVD) synthesis, and primary Investigation and data collection. Meryem B. performed the Formal Analysis of all characterization data and undertook the Writing – Original Draft Preparation of the manuscript.

**M. Nasuh ARIK:** Contributed to the Formal Analysis (specifically the XPS data interpretation) and participated in the Writing – Review & Editing processes of the manuscript.

**Ali Altuntepe:** Contributed to the experimental Investigation by performing the Raman spectroscopy characterization and carried out the necessary Validation steps.

**Hakan ATEŞ and Recep ZAN:** Provided Supervision for the entire doctoral research project and offered substantial intellectual contributions through the Writing – Review & Editing of the manuscript for scientific content and academic rigor.

**Acknowledgement**

This work was supported by The Scientific and Technological Research Council of Türkiye (TÜBİTAK) under the 1002-A Short-Term Support Module (Project No: 225M094).